\documentclass{PoS}

\usepackage{epsfig}

\title{Monte Carlo Tools for charged Higgs boson production}

\ShortTitle{MC Tools for charged Higgs}

\author{\speaker{Karol Kova\v{r}\'{\i}k}\\
        Institut f{\"u}r Theoretische Physik, Westf{\"a}lische Wilhelms-Universit{\"a}t M{\"u}nster,
	              Wilhelm-Klemm-Stra{\ss}e 9, D-48149 M{\"u}nster, Germany
		\\
        E-mail: \email{karol.kovarik@uni-muenster.de}}

\abstract{In this short review we discuss two implementations of the charged Higgs boson production process in association
with a top quark in Monte Carlo event generators at next-to-leading order in QCD. We introduce the MC@NLO and the POWHEG method
of matching next-to-leading order matrix elements with parton showers and compare both methods analyzing the charged Higgs boson
production process in association with a top quark. We shortly discuss the case of a light charged Higgs boson where the associated
charged Higgs production interferes with the charged Higgs production via $t\bar{t}$-production and subsequent decay of the top quark.}

\FullConference{Prospects for Charged Higgs Discovery at Colliders - CHARGED 2014,\\
		16-18 September 2014\\
		Uppsala University, Sweden}

\begin{document}

\section{Introduction}
Even after the discovery of the Higgs boson, the search still continues for particles beyond the Standard model (BSM).
A discovery of a charged Higgs boson would be one possible signal for many models beyond the Standard model with an extended Higgs sector.
Perhaps the simplest BSM model is the two-Higgs-doublet model (2HDM) which as the name indicates includes an additional Higgs doublet
with additional CP-even and CP-odd neutral Higgs bosons as well as a charged Higgs boson. Different types of 2HDM models differ in the 
way Higgs doublets couple to fermions. The most popular extension of the Standard model, the minimal supersymmetric standard model 
(MSSM), contains an extended Higgs sector with an additional Higgs doublet and a specific coupling to fermions identical to the type-II 
2HDM.

The most promising production channels for a charged Higgs boson depend on its mass and on the particular model. If, for example, the charged Higgs boson
is lighter than the top quark, then it is dominantly produced in decays of top quark $t\rightarrow H^+b$. If, on the other hand, it is heavier than the top quark,
the dominant production mechanism at a hadron collider is the direct production where the charged Higgs boson is produced in association with a top quark.

The next-to-leading order (NLO) prediction in perturbative QCD of the total cross-section of the direct production of a charged Higgs boson in association with a 
top quark has been known for some time \cite{Plehn:2002vy,Berger:2003sm}. The current state-of-the-art is the combination of NLO parton matrix elements with Monte Carlo event generators. 
This combination improves on the NLO predictions by using the parton shower algorithm of the Monte Carlo event generator to simulate the effects of further
soft and collinear enhanced radiation. There are several possible approaches to coupling NLO parton matrix elements with parton shower algorithms of 
Monte Carlo event generators. One such approach is the MC@NLO algorithm \cite{Weydert:2009vr} which couples the NLO charged Higgs production with the parton shower algorithm of the HERWIG event generator \cite{Corcella:2000bw}. In order to couple to different event generators and to improve on the negative weight events which arise in the MC@NLO framework, an alternative POWHEG \cite{Nason:2004rx,Frixione:2007vw,Alioli:2010xd} approach was devised which produces events with positive weight and can be coupled to any event generator such as HERWIG or PYTHIA \cite{Sjostrand:2006za}.

\section{NLO and Monte Carlo generators}

In recent years Monte Carlo generators reached a new level of precision where the new state-of-the-art is a Monte Carlo generator which
uses predictions from NLO matrix elements properly matched with parton showers. The matching is the crucial part of combining a NLO
prediction with a parton shower algorithm. Virtual parts of next-to-leading order matrix elements contain ultraviolet and infrared 
divergencies. The first can be removed by a suitable redefinition of parameters of the theory but in order to remove the infrared 
divergencies one has to include a process with one additional parton in the final state. Adding partons to the final state is also 
something the parton showers do and so it is natural that a conflict might arise. In order to understand how to consistently match
NLO matrix elements and parton showers, we have to look more closely at the form of a NLO cross-section. The NLO cross-section can be 
written as
\begin{equation}\label{nlom}
	\sigma = \int d\Phi_B\Big[B(\Phi_B)+\hat{V}(\Phi_B) + \int d\Phi_{\rm rad}\,C(\Phi_R(\Phi_B,\Phi_{\rm rad}))\Big] + \int d\Phi_R \Big[R(\Phi_R)-C(\Phi_R)\Big]
\end{equation}
Both the born contribution ($B$) and the virtual contribution ($V$) to the cross-section are integrated over the same phase-space $\Phi_B$ whereas the real contribution which is introduced in order to cancel the infrared divergence, is integrated over the phase-space with one additional parton ($\Phi_R$). The cancellation of the infrared divergence can elegantly be performed by introducing 
a counterterm $C$ which is analytically integrable over the additional one-particle phase-space $\Phi_{\rm rad}$ and has the same
divergent behavior as the matrix element. This counterterm in its integrated form cancels the divergence of the virtual contribution and in its unintegrated form cancels the divergence of the real matrix element. The Eq.~(\ref{nlom}) should be compared to the 
expression for a differential cross-section with at most one additional parton emission in the parton shower language
\begin{equation}\label{PSsig}
	d\sigma = d\Phi_B\, B(\Phi_B)\left(\Delta_i(t_I,t_0)+\sum_{(j,k)}\Delta_i(t_I,t)\frac{\alpha_s(t)}{2\pi}P_{i,jk}(z)
	\frac{dt}{t}dz\frac{d\phi}{2\pi}\right)\,,
\end{equation}
where $\Delta_i(t_I,t_0)$ is the Sudakov form factor which stands for no emission probability. Expanding this expression in the coupling constant $\alpha_s$, we obtain
\begin{equation}
	d\sigma = d\Phi_B\, B(\Phi_B)\left(1-\sum_{(j,k)}\int\frac{dt'}{t'}\int dz\frac{\alpha_s(t')}{2\pi}P_{i,jk}(z) + \sum_{(j,k)}\frac{\alpha_s(t)}{2\pi}P_{i,jk}(z)\frac{dt}{t}dz\frac{d\phi}{2\pi}\right)\,.
\end{equation}
The first/second term is the approximate substitute for the virtual/real corrections in the parton shower formalism and both of these
terms have to be dealt with during the matching process to prevent double counting. Here we will mention two different methods for matching NLO matrix elements with parton showers.
\paragraph{MC@NLO method}
In the MC@NLO method, the shower algorithm is left alone and instead the collinear approximation which is used in the shower, is also 
used as the collinear counterterm to cancel the collinear divergence in the NLO matrix element. In this case the extension of the  parton shower to include the NLO matrix element leads to a modification of the original parton shower expression in Eq.~(\ref{PSsig})
\begin{equation}\label{mcnlosig}
	d\sigma = d\Phi_B\, \bar{B}^{MC}(\Phi_B)\left(\Delta^{MC}(t_I,t_0)+\Delta^{MC}(t_I,t)\frac{R^{MC}(\Phi)}{B(\Phi_B)}\,d\Phi_{\rm rad}^{MC}\right)+\Big(R(\Phi)-R^{MC}(\Phi)\Big)d\Phi\,,
\end{equation}
where the modified born matrix element $\bar{B}^{MC}$ includes the virtual contribution and the integrated collinear counterterm
\begin{equation}
	\bar{B}^{MC}(\Phi_B) = B(\Phi_B) + \Big[V(\Phi_B)+\int R^{MC}(\Phi)\,d \Phi_{\rm rad}^{MC} \Big]\,.
\end{equation}
The last term in Eq.~(\ref{mcnlosig}) improves the description of the radiation of one additional parton by using the full matrix 
element. Using the collinear approximation of the shower as a collinear counterterm, makes this method dependent on the parton shower
algorithm one wants to match the NLO matrix element with.
\paragraph{POWHEG method}
An alternative approach is to modify the shower so that the first hardest emission is performed using the full matrix and the remaining
softer emissions are performed using the parton shower algorithm. This approach is called the POWHEG method and it avoids a possible double counting by clearly separating the first hardest emission from the remaining emissions. The cross-section in this approach looks
similar to the one in Eq.~(\ref{mcnlosig})
\begin{equation}
	d\sigma = d\Phi_B \bar{B}^S(\Phi_B)\left(\Delta^S_{t_0}+\Delta^S_t\frac{R^S(\Phi)}{B(\Phi_B)}d\Phi_{\rm rad}\right)+R^Fd\Phi_R\,,
\end{equation}
where the real contribution to the next-to-leading matrix element was split into the singular and finite part $R=R^S+R^F$ and the 
modified born matrix element again includes the virtual contribution
\begin{equation}
	\bar{B}^S(\Phi_B) = B(\Phi_B) + \Big[V(\Phi_B)+\int R^S(\Phi)\,d \Phi_{\rm rad}\Big]\,.
\end{equation}
The first hardest emission is excluded from the parton shower by using the vetoed showers and redefining the Sudakov form factor as
\begin{equation}
	\Delta^S_t = \exp\left[-\int\theta (t_r-t)\frac{R^S(\Phi_B,\Phi_{\rm rad})}{B(\Phi_B)}d\Phi_{\rm rad}\right]\,.
\end{equation}
The POWHEG method is independent of the shower algorithm and can be used with any Monte Carlo program provided it allows for vetoed 
showers.
\section{Charged Higgs production}
\begin{figure}[!h]
 \centering
 \epsfig{file=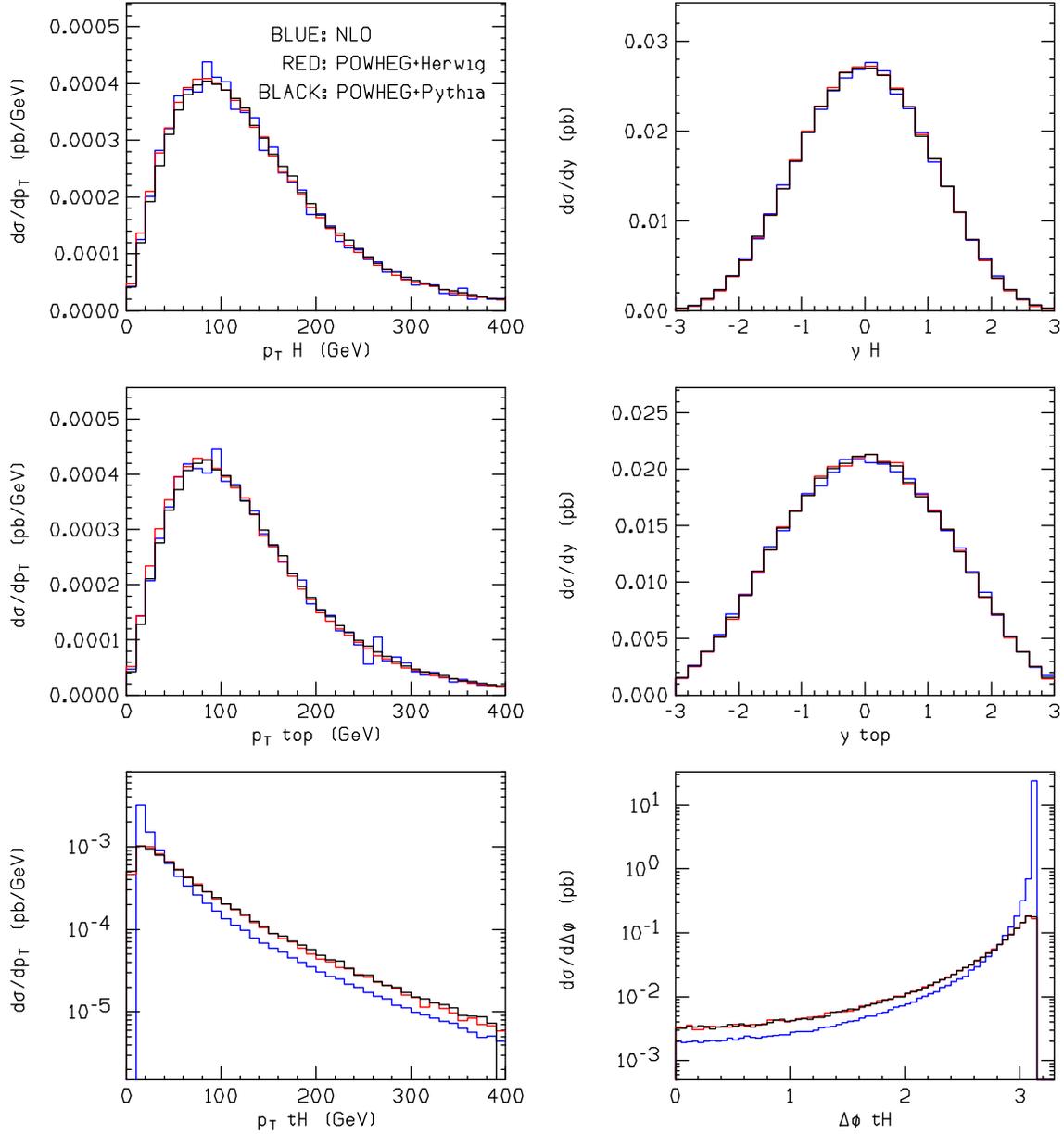,angle=90,width=\textwidth}
 \caption{\label{fig1}Distributions in transverse momentum $p_T$ (top left) and
 rapidity $y$ (top right) of the charged Higgs boson, $p_T$ (center left) and $y$
 (center right) of the top quark, as well as $p_T$ (bottom left) and azimuthal opening angle
 $\Delta\phi$ (bottom right) of the $tH^-$ system produced at the LHC with $\sqrt{s}
 =14$ TeV. We compare the NLO predictions without (blue) and with matching to
 the PYTHIA (black) and HERWIG (red) parton showers using POWHEG in the Type-II
 2HDM with $\tan\beta=10$ and $m_H=300$ GeV.}
\end{figure}

\begin{figure}[!h]
 \centering
 \epsfig{file=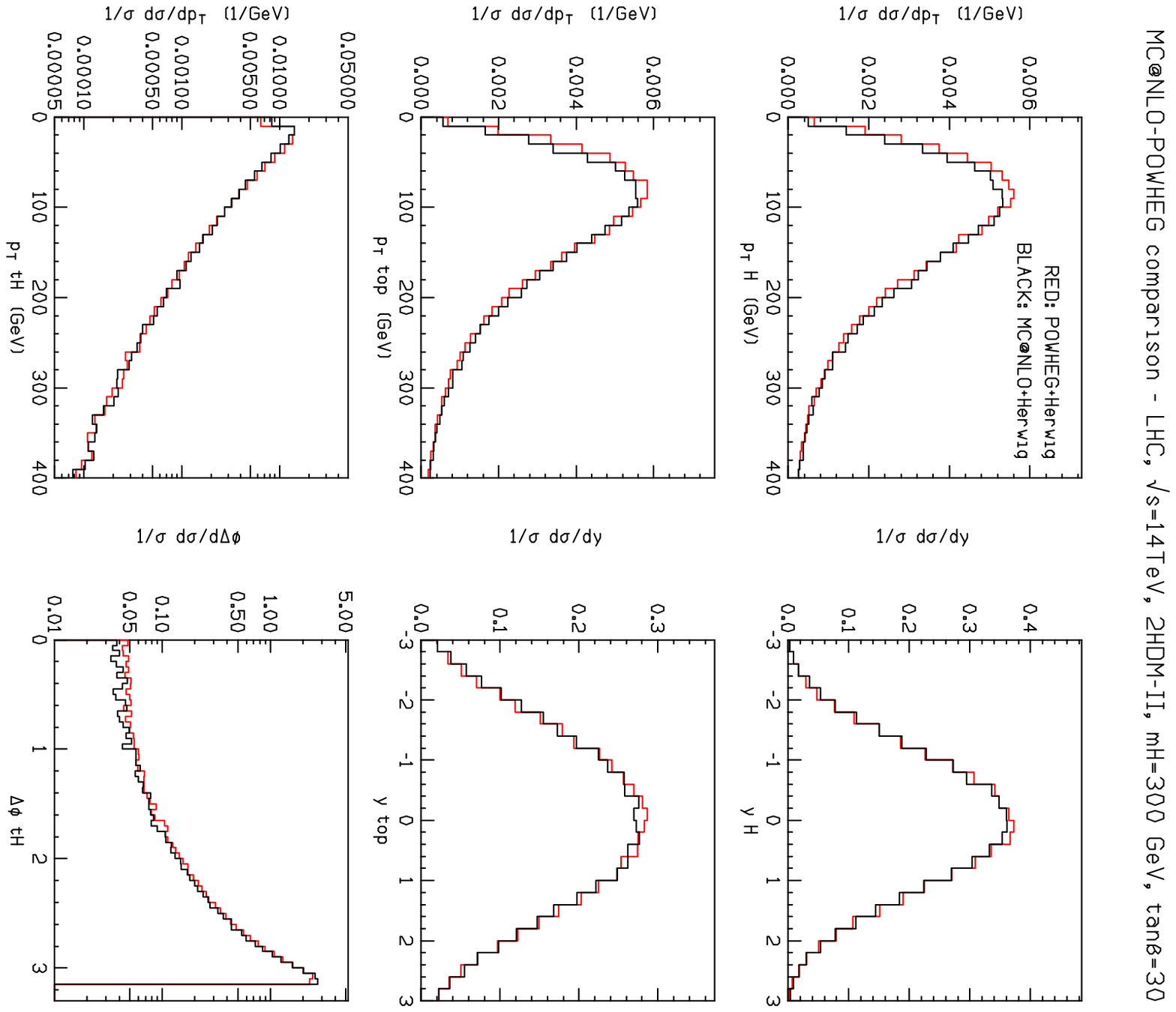,angle=90,width=\textwidth}
 \caption{\label{fig:2} We compare the charged Higgs boson production in the Type-II
 2HDM with $\tan\beta=30$ and $m_H=300$ GeV implemented in POWHEG and MC@NLO. We show the 
 same distributions as in Fig.\ \protect\ref{fig1}.}
\end{figure}

The charged Higgs production in association with the top-quark at the next-to-leading-order in QCD is currently implemented in both 
MC@NLO \cite{Weydert:2009vr} and POWHEG \cite{Klasen:2012wq} Monte Carlo event generators. In Fig.~\ref{fig1} we show a comparison of 
the implementation of the charge Higgs 
production using the POWHEG method matched to the Pythia and Herwig parton showers with the simple NLO prediction for heavy charged 
Higgs bosons. We see the effect 
of the radiation of multiple partons through the parton shower on the $p_T$-distribution and azimuthal opening angle distribution of the 
$tH^-$ system. A comparison of the implementations using the MC@NLO and POWHEG methods both matched with the Herwig parton showers
in Fig.\ \ref{fig:2} demonstrates the compatibility of both approaches.

\begin{figure}[!h]
 \centering
 \epsfig{file=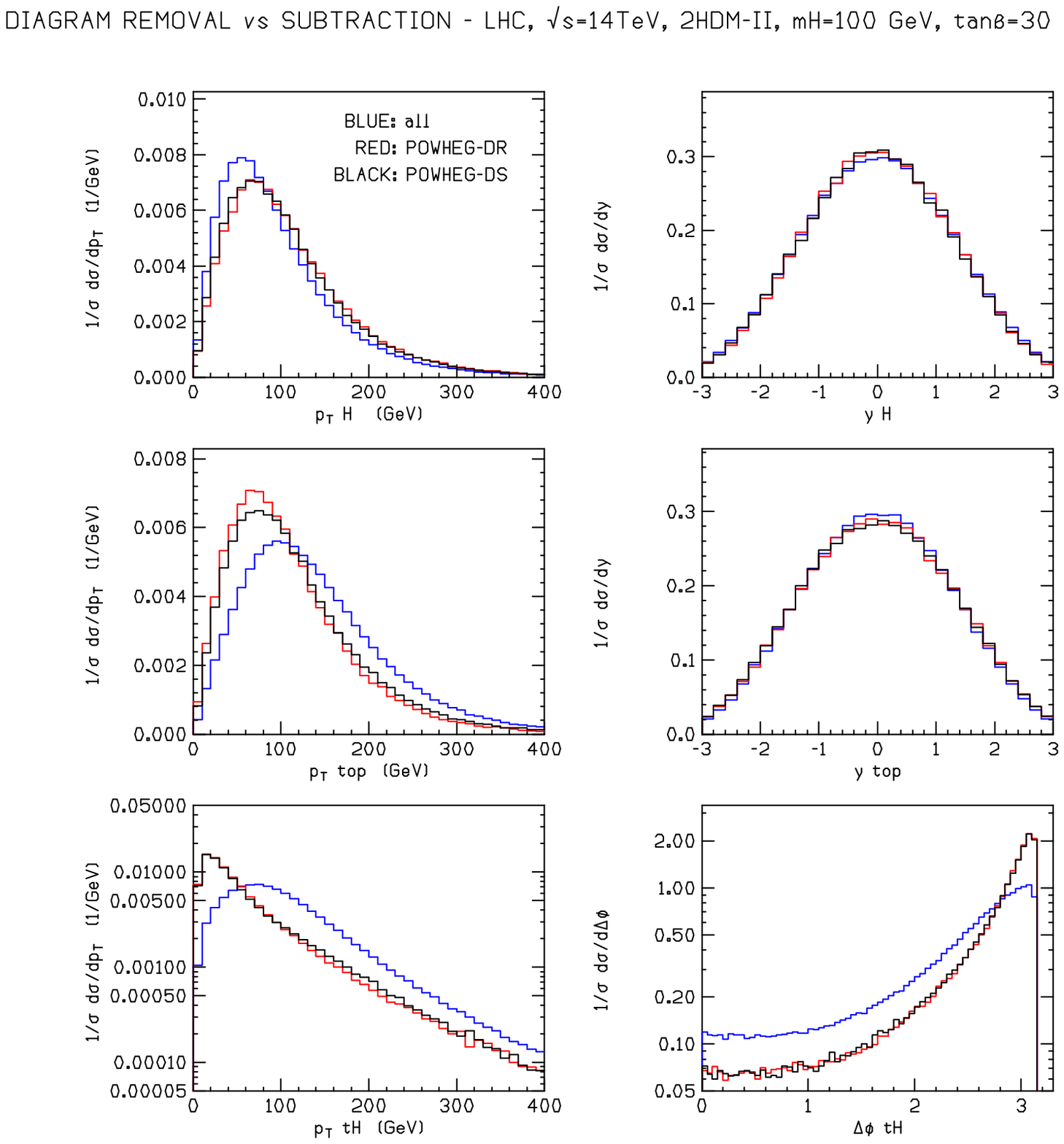,angle=90,width=\textwidth}
 \caption{\label{fig:3} We compare diagram removal and diagram subtraction method to isolate 
 the associated charged Higgs boson production in the Type-II
 2HDM with $\tan\beta=30$ and $m_H=100$ GeV implemented in POWHEG. We show the 
 same distributions as in Fig.\ \protect\ref{fig1}.}
\end{figure}

\begin{figure}[!h]
 \centering
 \epsfig{file=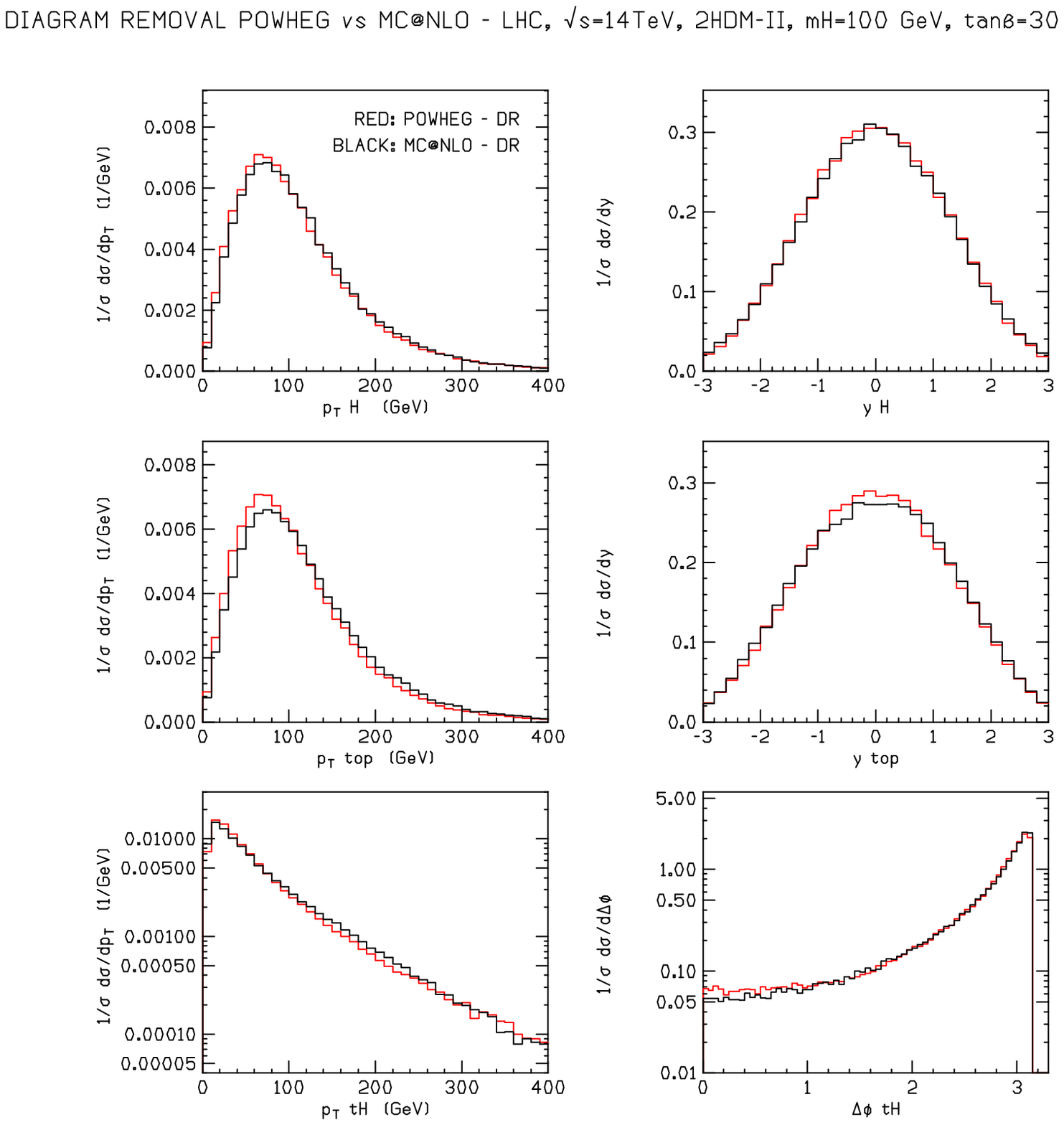,angle=90,width=\textwidth}
 \caption{\label{fig:4} We compare diagram removal method to isolate 
 the associated charged Higgs boson production in the Type-II
 2HDM with $\tan\beta=30$ and $m_H=100$ GeV implemented in POWHEG and MC@NLO. We show the 
 same distributions as in Fig.\ \protect\ref{fig1}.}
\end{figure}

\begin{figure}[!h]
 \centering
 \epsfig{file=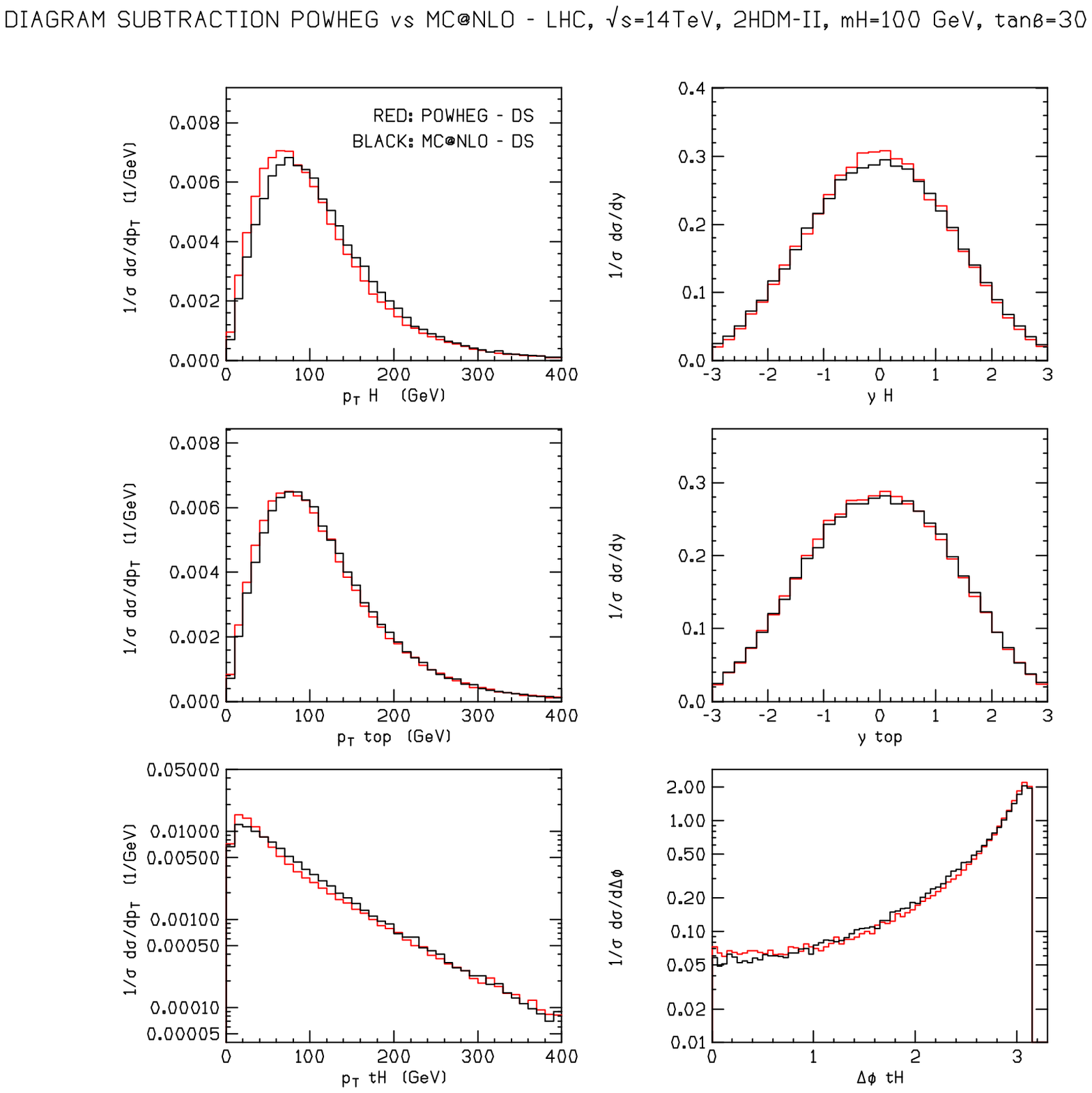,angle=90,width=\textwidth}
 \caption{\label{fig:5} We compare diagram subtraction method to isolate 
 the associated charged Higgs boson production in the Type-II
 2HDM with $\tan\beta=30$ and $m_H=100$ GeV implemented in POWHEG and MC@NLO. We show the 
 same distributions as in Fig.\ \protect\ref{fig1}.}
\end{figure}

As already extensively discussed in \cite{Weydert:2009vr,Klasen:2012wq,Frixione:2008yi}, the case where the charged Higgs boson is lighter than the 
top-quark, has to be handled with care. In this case the leading production mechanism of the charged Higgs boson at the LHC is through 
the decay of a top quark which is produced alongside an anti-top quark. At leading order the production of the charged Higgs boson 
in association with a top quark and the production of charged Higgs boson via $t\bar{t}$-production and a subsequent decay are independent. At next-to-leading order though, these processes cannot be separated and an interference between them arises.
One would nevertheless like to separate the two production processes so that they can later be joined but with the $t\bar{t}$-production
generated separately using NLO precision. 
\begin{equation}
	 \vert \mathcal{M}_{ab} \vert^2  =  \vert \mathcal{M}_{ab}^{tH^-} \vert^2 + 2 {\rm Re} \bigl(\mathcal{M}_{ab}^{tH^-} \mathcal{M}_{ab}^{t \bar{t}*} \bigr) +  \vert \mathcal{M}_{ab}^{t \bar{t}} \vert^2
	= \mathcal{S}_{ab} + \mathcal{I}_{ab} + \mathcal{D}_{ab}
\end{equation}
There were two methods put forward in \cite{Frixione:2008yi}. In the first method called diagram removal one removes all resonant $2\rightarrow 3$ diagrams which belong to the $t\bar{t}$-production from the associated production. When removing the diagrams at amplitude level, one looses also any information on the interference between the processes ($\mathcal{I}_{ab}$).
The second option called diagram subtraction subtracts the resonant $t\bar{t}$-contribution from the cross-section by subtracting
\begin{equation}
	d \sigma^{\rm sub}_{H^-t}=\frac{f_{\rm BW}(m_{H^-\bar{b}})}{f_{\rm BW}(m_t)}
	\left|\tilde{\cal A}^{(t\bar{t})}\right|^2\,.	
\end{equation}
This procedure leaves the interference effects present in the predictions for the associated production. We compare both methods which are implemented in MC@NLO and POWHEG. In Fig.\ \ref{fig:3} we compare both methods implemented in POWHEG matched to the Herwig parton showers and in Figs.\ \ref{fig:4}-\ref{fig:5} we compare the implementations of the methods in POWHEG and MC@NLO. 
\section{Conclusion}
We have provided a short review of the current implementation of the charged Higgs boson production in association with the top-quark
for heavy and light Higgs boson in the 2HDM. We have shown that both implementations in POWHEG and MC@NLO discussed here are in excellent agreement for both heavy and light Higgs bosons.

\end{document}